\documentclass[sigconf]{acmart}
\usepackage{setspace}

\usepackage[english]{babel}
\usepackage{makecell}
\usepackage{float}

\usepackage{graphicx}
\graphicspath{{figures/}}
\usepackage{balance}
\usepackage{url}
\usepackage[hide]{mnotes}
\Mnewauthor[SL]{SL}{olive}
\Mnewauthor[MM]{MM}{orange}
\makeatletter
\long\def\put(#1,#2)#3{%
  \@killglue
  \@defaultunitsset\@tempdimc{#2}\unitlength
  \raise\@tempdimc
  \hbox to\z@{
    \@defaultunitsset\@tempdimc{#1}\unitlength
    \kern\@tempdimc
    #3\hss}%
  \ignorespaces}
\makeatletter  

\acmConference[MAPII-2022]{}{June 07,  2022}{Rome}
\begin{document}

\title{Temporal and Spatial Elements in \\Interactive Epidemiological Maps}

\author{Saturnino Luz}
\email{s.luz@ed.ac.uk}
 \orcid{0000-0001-8430-7875}
\affiliation{%
  \institution{Usher Institute, Edinburgh Medical School}
  \institution{The University of Edinburgh}
\city{Edinburgh}
  \country{United Kingdom}
}
 \author{Masood Masoodian}
\email{masood.masoodian@aalto.fi}
\orcid{0000-0003-3861-6321}
\affiliation{%
  \institution{School of Arts, Design and Architecture}
  \institution{Aalto University}
	\city{Espoo}
  \country{Finland}
  }

\setcopyright{none}
\settopmatter{printacmref=false} 
\renewcommand\footnotetextcopyrightpermission[1]{} 
\pagestyle{plain}


\begin{abstract} 
Maps have played an important role in epidemiology and public health since the beginnings of these disciplines. With the advent of geographical information systems and advanced information visualization techniques, interactive maps have become essential tools for the analysis of geographical patterns of disease incidence and prevalence, as well as communication of public health knowledge, as dramatically illustrated by the proliferation of web-based maps and disease surveillance ``dashboards'' during the COVID-19 pandemic. While such interactive maps are usually effective in supporting static spatial analysis, support for spatial epidemiological visualization and modelling involving distributed and dynamic data sources, and  support for analysis of temporal aspects of disease spread have proved more challenging. Combining these two aspects can be crucial in applications of interactive maps in epidemiology and public health work. In this paper, we discuss these issues in the context of support for disease surveillance in remote regions, including tools for distributed data collection, simulation and analysis, and enabling  multidisciplinary collaboration. 
\end{abstract}




\maketitle

\section{Introduction}

The use of interactive maps and map-based interfaces in support of analytical tasks in epidemiology and public health has become increasingly widespread in recent years, having reached unprecedented levels during the ongoing COVID-19 pandemic. Epidemiology can be defined as ``the study of the distribution and determinants of health-related states or events in specified populations, and the application of this study to the prevention and control of health problems'' \cite{bib:Last01d}. This definition carries geographical, temporal and causal elements which have been the focus of map-based interfaces and systems, in public health and other domains.  The nature of the challenges faced by designers of such systems illustrates fundamental issues in spatio-temporal visualization. Tackling these issues involves addressing how map-based interfaces can support complex real-world scenarios and use cases involving collaborating multidisciplinary teams, as well as challenges related to the integration and appropriate granularity of spatial public health data, clustering identification and statistical modelling, visualization of change, and identification of causal relationships between variables that characterise the evolution of spatio-temporal events,

In this paper, we review these issues, and illustrate approaches to addressing them in interactive systems through the analysis of uses of map-based interfaces for supporting epidemiological research and decision-making in public health in collaborative settings. We start with a brief overview of traditional cartographic maps in general, and then introduce interactive maps and map-based interfaces, followed by their applications in epidemiology, and pointing out current developments and existing challenges in this area. Finally, we illustrate these issues by describing a practical application of map-based interfaces in research  on health, environmental and socioeconomic factors affecting the spread of infectious diseases in an Amazonian region, in South America, involving multidisciplinary collaboration.

\section{Maps}
While everyone would claim they know what a ``map'' is, the term \emph{map} itself may mean rather different things to different people, and  it may vary from person to person and culture to culture \cite{bib:Tyner2014a}, as a compilation of three hundred and twenty one different definitions of the word ``map'' by Andrews \cite{bib:Andrews1998} would attest. 

Similarly, although most people would think of a map as some form of representation of a geographical area, others would argue that maps can be used to represent many other kinds of information, some of which may even have nothing to do with physical reality -- for instance representing political, cultural, or historical information \cite{bib:KoponenAndHilden}. 

However, what all maps have in common is that they generalize and simplify the information they represent, rather that trying to fully represent any reality \cite{bib:Barber2005}, and by doing so, they may distort or compromise that information \cite{bib:AirikkaAndMasoodian2019}. All maps also project salient and relevant aspects of such information onto  a geometry --in the form of bounded objects -- with certain topological properties, in a manner to to geographical projections but not always corresponding to geographical objects.
As such, every map is a representation of reality and ``not reality itself'' \cite{bib:LambertAndZanin2020}.
Despite this, to function well, maps need to maintain a sufficient level of accuracy necessary for the tasks they are designed to support \cite{bib:AirikkaAndMasoodian2019}.  In addition, maps that more closely represent reality are more likely to support their intended tasks than those that do not  \cite{bib:MacEachren2004}. 

In that sense, geographical maps are perhaps the most functional type of maps, because they represent some form of physical reality, projected onto a 2-dimensional (2D) space \cite{bib:Tyner2014a}.  Indeed the earliest maps are thought to have been created to help people find their way  \cite{bib:Turchi2007} 
-- what is referred to more commonly as \emph{wayfinding} -- in some ``real'' physical or metaphysical world. There are of course many types of such geographical maps, each supporting a particular range of tasks.   These include, for instance, cartographic maps \cite{bib:KraakAndOrmeling2011}, tourist maps \cite{bib:JancewiczAndBorowicz2017,bib:AirikkaAndMasoodian2019}, road maps \cite{bib:Tyner2014a}, and public transport maps, such as the well-known London Underground map \cite{bib:Spence2014}.

Tyner \cite{bib:Tyner14pmd} focuses on the functionality of geographical maps, and divides them into three main categories:

\begin{itemize}
\item \emph{general-purpose maps} or \emph{reference maps} do not
  emphasise any particular geographical feature over another, but
  instead show the location of different geographic phenomena, such as
  cities, roads, rivers, etc.
\item \emph{special-purpose maps} are targeted at specific users and
  their needs, such as geological maps, soil maps, weather maps, etc.
\item \emph{thematic maps} tend to show a single distribution over a
  spatial background or framework, to help locating the distribution
  being mapped, such as population density, land use, family income,
  rain fall, etc.
\end{itemize}

In contrast, Lambert and Zanin \cite{bib:LambertAndZanin2020} focus more on representational aspect of geographical maps,  and divides them into tow main categories:

\begin{itemize}
\item \emph{topographic maps} represent concrete elements resulting
  from direct observations of, for instance, roads, waterways,
  buildings, etc.
\item \emph{thematic maps} represent localisable qualitative and
  quantitative information using rules of graphical
  semiology. \end{itemize}

Maps used in epidemiology and public health work are often of thematic type, regardless of whether their functional or representational aspects are considered first. Maps used for public health-related purposes -- in the form of disease maps \cite{bib:Koch2011} -- might constitute one of the earliest forms of thematic maps.  The use of such thematic maps to monitor, explain and predict patterns of disease spread as part of epidemiological reasoning dates back at least to the late 18th century \cite{bib:Koch2011}. The best known example of this type of disease map use case  is for that of the cholera epidemic in London in 1845, in which John Snow used a map to locate and show that the majority of cholera deaths occurred near  a water pump in Broad Street neighbourhood of Soho, and therefore propose that cholera is likely to be a water-borne disease \cite{bib:Tufte1990,bib:Friendly2008,bib:Koch2011}.

\section{Interactive Maps}

Regardless of their type, all maps are visualizations, and as such, they aim to support seeing and exploring their underlying data in different ways  \cite{bib:Tyner2014a}. More specifically, maps are designed to help their readers perform a range of tasks, including for instance  identifying locations of interest, getting data related to those locations, identifying any data patterns related to those locations, or making spatial comparison between such patterns at different locations on the same map, or between different maps \cite{bib:SlocumEtAl2008}. Sometimes it is also necessary to compare data patterns across time, either for the same location -- i.e., temporal comparisons -- or between different locations -- i.e., spatio-temporal comparisons.

Bertin \cite{bib:Bertin2010}, for instance, discusses how multiple maps juxtaposed on a 2D plane can help viewers to identify patterns in geospatial data more readily. Similarly, Carr and Pickle \cite{bib:CarrPickle2010} use the idea of \emph{small multiples}, as proposed by \cite{bib:Bertin2010}, to suggest \emph{micromaps} as a technique for highlighting geographical patterns in data.

While it is possible -- as has been the case for centuries -- to perform a wide range of tasks using static maps, it is clearly more effective and often more efficient to perform them using interactive maps \cite{bib:Tyner14pmd}. For instance, it has been argued that making side-by-side comparison of maps is not generally very effective for investigating spatial correspondence between them \cite{bib:MacEachren2004}. It has, therefore, been necessary to develop interactive techniques such as \emph{layering} to deal with some of the limitations of static maps \cite{bib:LuzAndMasoodian2014}. For example, Adrienko et al. review interactive methods for spatio-temporal visualizations \cite{bib:AndrienkoAndrienkoGatalsky03ex} and identify some of the challenges in the field \cite{bib:AndrienkoAndrienkoEtAl11c}.

Interactivity can, of course, take place at different levels and does depend on the application area and the context of use. It can refer to any tool, in the form of a standalone application or a webmap \cite{bib:Tyner2014a},  that  allows users to set different parameters used to generate, control, and display maps during their use, and by doing so, extend the user's interaction capabilities far beyond those possible with static maps \cite{bib:RistAndMasoodian2022}. Furthermore, while there are a wide range of interactive map systems, their level of interactivity is always constrained by the number of parameters they allow their users to manipulate, and to what extend and by which means  \cite{bib:RistAndMasoodian2022}. Despite such constraints, however, in many modern computer applications, maps go far beyond their traditional functionality of simply presenting data, and can therefore be considered as versatile interfaces to geospatial data \cite{bib:Kraak2006} as well as other types of data, including epidemiological and public health data. Such interactive map-based interfaces need to support ``information exploration and knowledge construction [...] without hypotheses about the data''  and through ``unencumbered search for structures and trends''  \cite{bib:MacEachrenKraak2001}.   In more recent years, the availability of powerful API for web-based map systems has made it possible to overlay geospatial data on interactive maps in a wide range of applications \cite{bib:LuzAndMasoodian2014}.  This is achieved using \emph{layers} of data, which can be turned on and off to show or hide different data sets that the users are familiar with \cite{bib:EliasEtal2008}.

\section{Interactive maps in epidemiology}

Static maps often function as {\em coordinative artifacts} \cite{bib:bardrambossen05awca,bib:SchmidtSimonee96coord} in enabling the {\em articulation work} \cite{bib:StraussFagerhaughEtAl17s} of multidisciplinary teams engaged in disease surveillance and epidemiological tasks. Maps have been used to plan interventions, to divide and coordinate the work of medical professionals and researchers, to track the progression of epidemics, and to focus discussion and analysis in co-located meetings. Effective coordination mechanisms can in fact be employed on shared physical maps with the help of physical actions (e.g., placement of push pins, annotation, etc) and careful design \cite{bib:Monmonier90s}. However, the advent of interactive maps has enabled vastly enhanced support for collaboration in epidemiology, as well as in other areas. Interactive, web-based ``dashboards'', for instance, have contributed to geovisualization tasks and made it possible for researchers, public health officials and policy makers to share up-to-date information on disease spread \cite{bib:Crisan22imdvc}. Interactive maps offer a powerful tool for the visualization of variations in disease burden in populations across places and time -- in what is known as {\em spatial epidemiology} -- helping characterize geographically global and local determinants of population health heterogeneity \cite{bib:EberthKramerEtAl21w,bib:MelikerSloan11s}.

As noted above, spatial elements have been an integral part of epidemiological work since the establishment of the discipline in its current form, supporting the key analytical tasks of discovering disease clusters, predicting disease spread, monitoring exposures, analysing location-related social determinants of health (such as environmental changes, neighbourhood infrastructure and socioeconomic demographics), and assessing the effects of public health interventions. Lately, technologies that enable large-scale data collection, including but not limited to crowd-sourcing, social media analysis, citizen science, and mobile collaborative tools have started to enjoy more widespread use in spatial epidemiology \cite{bib:WeiIbrahimEtAl20s}. While these tools have enjoyed popularity in the visualization and data science community and their potential has been acknowledged in spatial epidemiology, concerns remain about incompleteness, inconsistency and bias issues that often affect data acquired through these tools \cite{bib:EberthKramerEtAl21w}. Therefore, the need for more robust methods for aggregation of diverse data sources for analysis and visualization through interactive maps is well established \cite{bib:CarrollAuEtAl14v}. In response to this need, conventional epidemiological models have been supplemented --- and in some cases replaced by --- new methods such as agent-based modelling \cite{bib:EubankGucluEtAl04m} to facilitate the incorporation of molecular epidemiological \cite{bib:CarrollAuEtAl14v} and social network data into map representations \cite{bib:LuzMasoodian22inbaps}.

In addition to the challenges related to the incorporation of new technologies and data sources to epidemiological maps, other more fundamental issues remain, some of which are exacerbated by the greater availability of data that has followed the introduction of web and mobile communications technology, and by the nature of interactive visualizations. These issues can be classified into two main groups: inherent constraints of spatial representation and issues representing temporal information.

\subsection{Spatial Representation}
As regards spatial representation, challenges include: the risk of cognitively overloading or misleading users \cite{bib:CarrollAuEtAl14v}, the granularity of geo-referenced data \cite{bib:EberthKramerEtAl21w}, security and privacy issues, and the need to integrate map visualizations within different public health tasks \cite{bib:LuzMasoodianCesarioBIT15,bib:MasoodianLuzKavenga16n} and epidemiological processes and patterns  \cite{bib:Jacquez00s}. 

Map representations of disease spread may mislead users by suggesting the presence of visual patterns or clusters which turn out to be spurious products of spatial autocorrelation -- such as the fact that values for one geographic location naturally tend to be similar to those of nearby locations -- or conversely, by obscuring the presence of actual clusters due to arbitrary boundaries (e.g., country borders). 

Fortunately, statistical methods for analysis of autocorrelation have been proposed which help users assess potential distortions due to spatial proximity \cite{bib:EberthKramerEtAl21w}, and kernel methods have been used to overcome issues relating to the impact of geographical boundaries (e.g., see \cite{bib:BurkeHeft-NealBendavid16ssaf}). 

The issues of geographical granularity and privacy are inter-related. While geo-referenced epidemiological data are usually available in aggregated form, there is arguably a need for more granular data, at the level of individual position and mobility, such as the information that can be derived from the GPS devices which are now ubiquitous thanks to the penetration of smartphone technology. Such information could be used, for instance, to determine individual exposure and predict outbreaks. However, reconciling the needs of public health and individual rights can be a delicate balancing act, as illustrated by experiences of some countries during the COVID-19 pandemic. 

Finally, there are many different uses for static and interactive maps, and users of spatial epidemiological data have likewise diverse tasks to perform, often involving different sets of requirements. We review an example of such diversity in more detail in Section~\ref{sec:an-exampl-inter}.

\subsection{Temporal Representation}
Integrating temporal aspects to essentially spatial artefacts such as maps poses challenges that go beyond those faced in epidemiological contexts \cite{bib:AndrienkoAndrienkoEtAl11c,bib:AndrienkoAndrienkoGatalsky03ex}. Time is of vital importance to epidemiology, as the discipline seeks to identify and test causal relationships between exposures and outcomes, and time is intrinsic to causal analysis.  

Spatio-temporal visualization can be accomplished in static maps through multiple snapshots depicting cross-sections of the temporal evolution of events of interest (e.g., successive depictions of infection rates over a geographical area) or location changes \cite{bib:Renolen97trr,bib:Langran20tgins}. However, interactive maps can also utilise animation, in the form of snapshots in time, movement histories and time windows \cite{bib:AndrienkoAndrienkoGatalsky03ex}, to reveal change patterns that might not otherwise be apparent in static snapshots. Such designs can be conceptualised as treating time as a cartographic variable -- similar to Bertin's visual variables \cite{bib:Bertin2010} -- as suggested by cartographic research \cite{bib:Monmonier90s,bib:MacEachren2004}. 

As with spatial epidemiology, temporal epidemiological maps face a need for underlying disease and process models that can be better integrated to spatio-temporal analysis \cite{bib:MelikerSloan11s}. In this regard, and as we have pointed out, agent-based models \cite{bib:EubankGucluEtAl04m} which can better account for complex boundary conditions and the spatio-temporal dynamics of disease spread, offer a promising alternative to traditional differential equation models \cite{bib:LuzMasoodian22inbaps}. 

Finally, temporal map-based visualizations need better support for causal inference, to enable the user to relate spatio-temporal changes to statistical data, and make assumptions about causal relationships \cite{bib:Pearl00causal}. This is a complex issue that will likely provide fertile ground for research for many years to come.

\section{An example: interactive maps in disease surveillance research}
\label{sec:an-exampl-inter}

A few years ago, we (the authors) participated in a multidisciplinary research project aimed at studying the determining factors in the spread of neglected infectious diseases in the Peru-Bolivia-Brazil ``tri-national'' borders \cite{bib:CesarioCesarioAndrade-Morraye11enam}. This study encompassed complex, inter-related environmental, geographical and socioeconomic factors, including global climate change, the dynamics of land use and cover in the tropical forest, migration patterns, population dispersion, and access to healthcare. Therefore, to support the computational and analytical requirements of that team, we approached health management from a perspective that combined human ecology, disease surveillance, and patient care.

As the team comprised researchers (epidemiologists, human ecologists) and healthcare professionals (clinicians and public health managers), we adopted a broad perspective of information exchange, whereby support for collaboration needs could be provided at different levels, in aid of different though inter-related activities, covering support for nurses and community healthcare workers working in remote locations, tools for collection and maintenance of patient records, and provision of exposure and disease data to epidemiological surveillance bodies. 

To this end, we developed a set of tools that combined mobile devices for patient care and epidemiology research in the field, centralised databases for modelling and generation of alerts, and support for both synchronous and asynchronous communication. These tools were tied together through map-based interfaces which served as the team's basic coordination mechanisms \cite{bib:SchmidtSimonee96coord}, while retaining the ability to support tasks performed by specific groups. 

In this section, we will examine two specific uses of interactive maps in that project to illustrate this approach. These are mobile support for fieldworkers in collecting and monitoring of disease data \cite{bib:LuzMasoodianCesarioBIT15}, and collaborative analysis of spatial epidemiological data by researchers and public health managers \cite{bib:MasoodianLuzKavenga16n}.

\subsection{Mobile Support for Fieldworkers}
To support data collection by health fieldworkers, we developed a prototype implemented on a tablet computer which integrated collection of patient data, assistance for detection of common and monitored diseases, communication with specialists from regional hospitals and central data repositories, and spatio-temporal visualization of case reporting. 

\begin{figure}[!b]
  \centering
  \includegraphics[width=\linewidth]{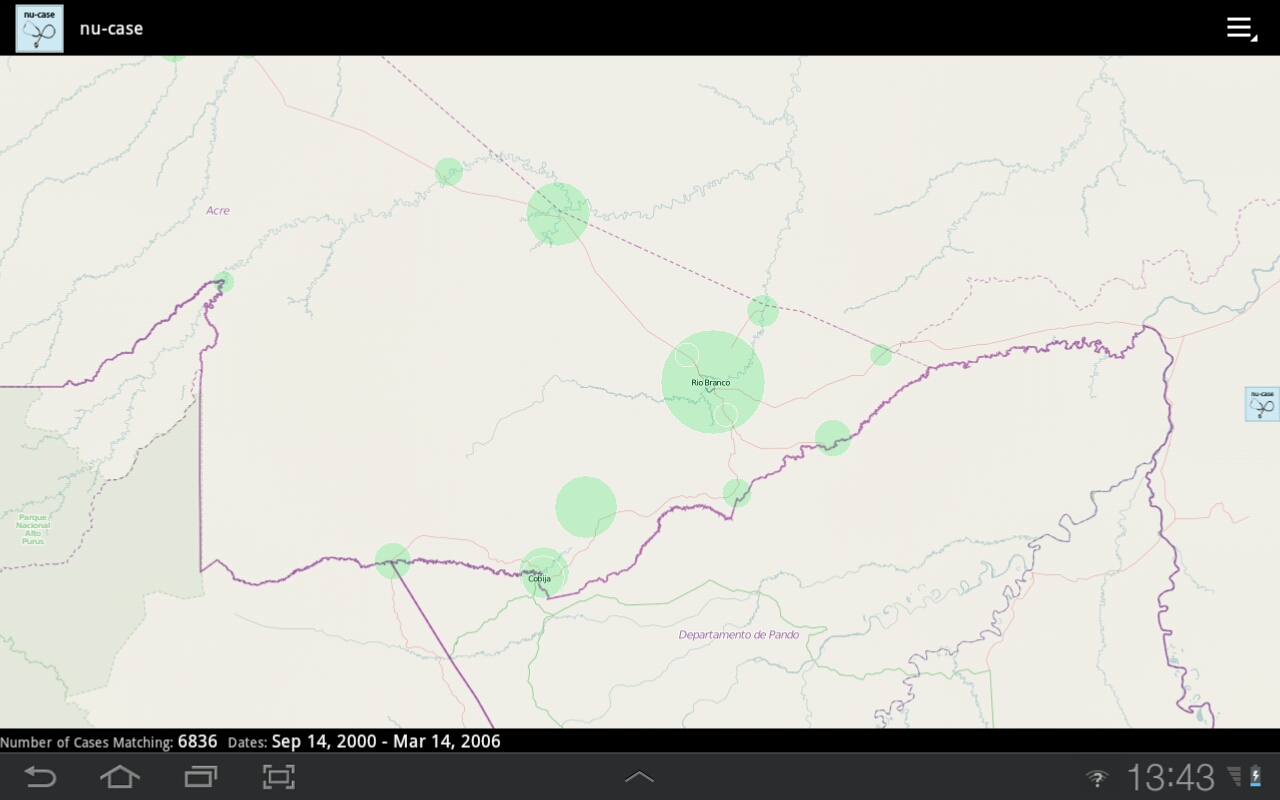}
  \caption{Map-based visualization of case incidence reports. Circles represent number of cases in different settlements, with their diameters being proportional to the number of reported cases.}
  \label{fig:nucasevismap}
\end{figure}

Data collection was enabled through speech and touch input modalities, with support for the recording of GPS coordinates.  Patient data were collected locally and aggregated into a central repository which could then be accessed remotely for visualization on spatial and temporal dimensions. The system displayed case occurrence and distribution on a map of the region. Figure~\ref{fig:nucasevismap} depicts the map-based interface of this tool. 

\begin{figure}[!t]
  \centering
  \includegraphics[width=\linewidth]{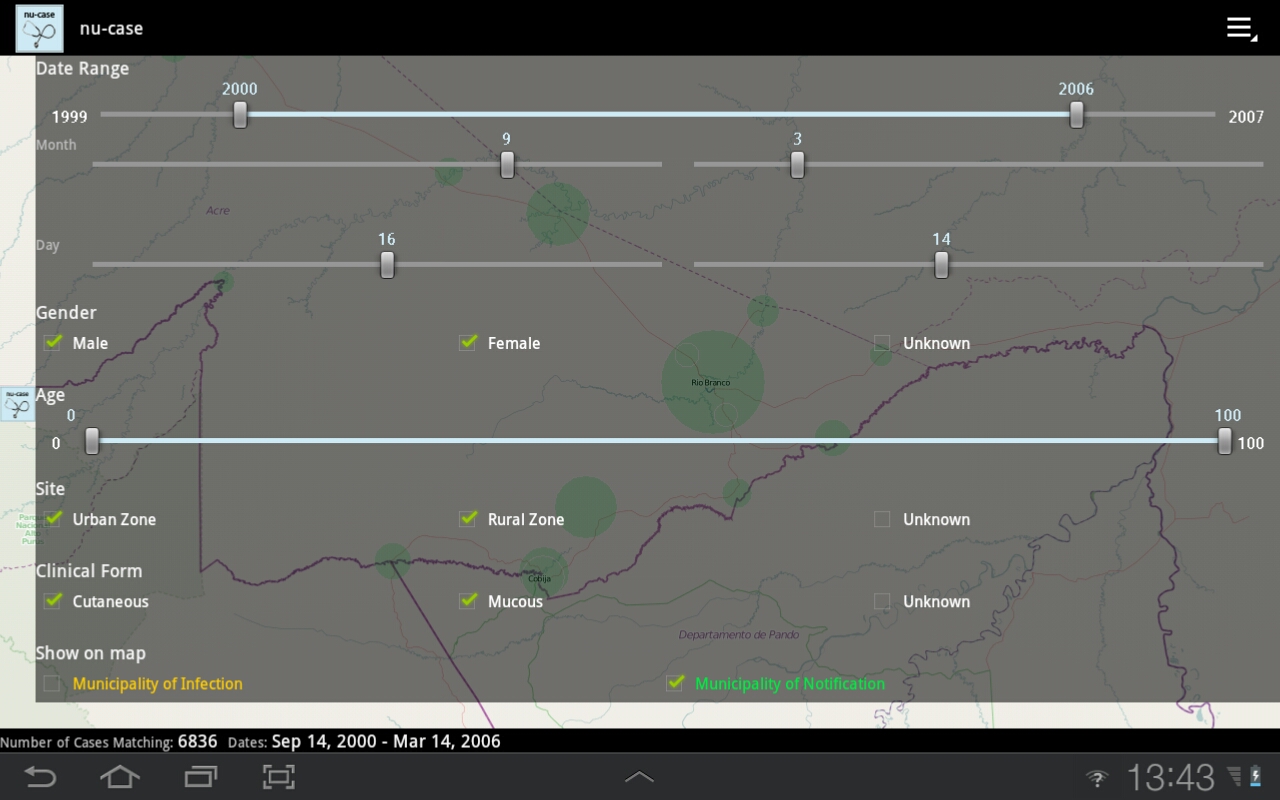}
  \caption{Query interface with support for identification of spatio-temporal patterns of disease spread through use of animation techniques.}
  \label{fig:nucasevisquery}
\end{figure}

Cases could be displayed by municipality of infection or municipality of notification, to help identify possible disparities between the project's and official data collection and notification processes. Circles with diameters proportional to the number of cases in an area are shown on the map. The user was able to alter the display by selecting different date ranges and combinations of features from patient records, as shown in Figure~\ref{fig:nucasevisquery}. Animation was employed to display case number progression over time -- for different spatial parameters settings -- to enable trend analysis and identification of disease spread patterns.

The data collection map-based interface was evaluated positively by the different user groups. A questionnaire administered to 47 participants (15 local doctors, 15 epidemiologists, 13 nurses and 4 public health administrators), found that users were very positive in rating the system's functionality on a 7-point Likert scale, specially in relation to recording geographical data (GPS) (mean 6.4, $sd=1.1$), accessing patient cases  (6.3, $sd=1.1$), and visualising geographical distribution of case data (6.6, $sd=0.8$). Disease monitoring was ranked  as the most important task supported by the system, followed by support for medical research and case notification and records. Participants pointed out that medical research and disease surveillance are inter-related tasks which involve a number of information seeking and data collection sub-tasks which could be facilitated by networked mobile devices deployed in the field.

\subsection{Support for Collaborative Data Analysis}
To support collaborative data analysis by researchers, we implemented a system that enabled visualization of epidemiological data on a large display with which users could interact through their personal mobile tablet devices so as to be able to share specific data and analyses with the group. Visual communication at meetings through sharing of data on large displays is typical of multidisciplinary medical teamwork. In such settings, where different experts present and discuss evidence from a variety of sources, a shared display usually acts as focus for discussion and as a means for establishing and recording the team's common ground on the case under discussion \cite{bib:LuzTOIS12}. Thus while our first use case focused on map-based interaction for data collection and epidemiological modelling in a distributed setting where communication was mostly asynchronous, this second use case focused on synchronous, co-located collaboration \cite{bib:MasoodianLuzKavenga16n}.

The system supported coupled and decoupled modes across an ensemble consisting of a large shared display and individual mobile device screens. Coupled interactions -- involving networked use of large and small displays at the same time -- took place around group visualizations, while decoupled interactions took place on private small displays. The system enabled sharing of personal visualizations in the context of the group visualization. Using these combinations, the system allowed shifting between visual analytics processes carried out individually on personal visualizations and shared activities carried out together on a group visualization.

\begin{figure*}[tbhp]
\centering
\parbox{.5\linewidth}{
\includegraphics[width=0.99\linewidth]{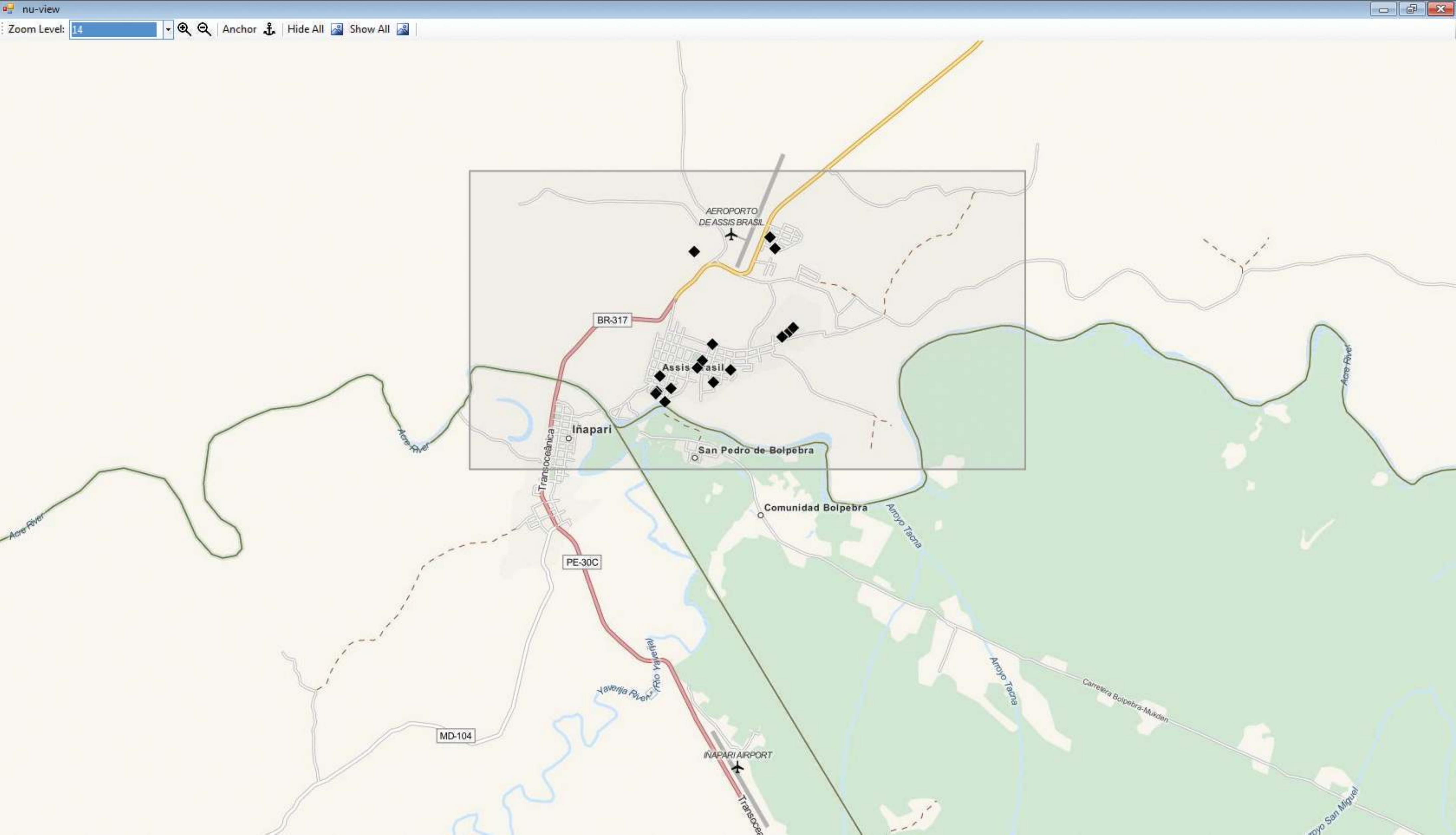} a)}%
\parbox{.5\linewidth}{
\includegraphics[width=0.99\linewidth]{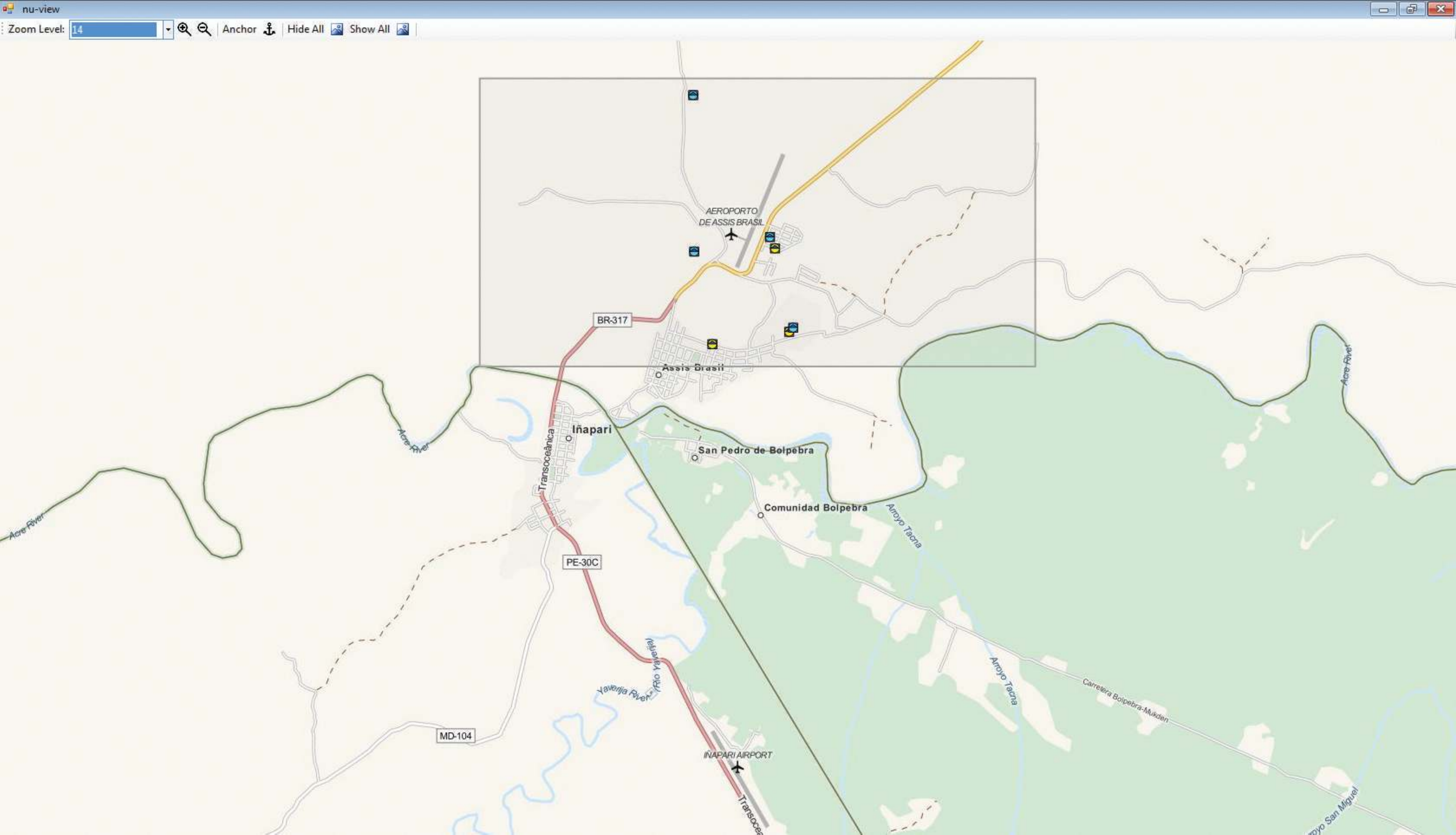} b)}
\parbox{.5\linewidth}{
\includegraphics[width=0.99\linewidth]{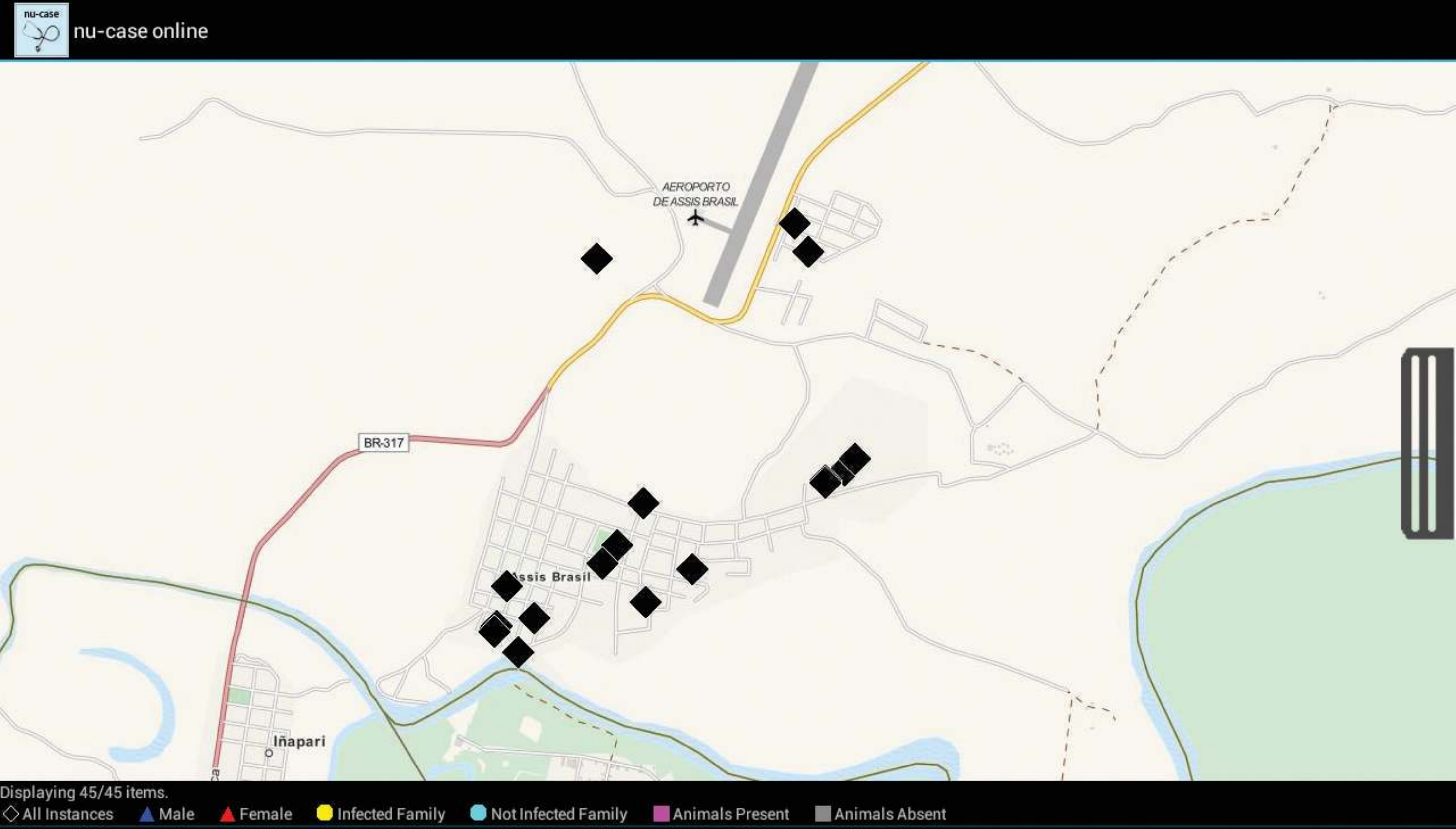} c)}%
\parbox{.5\linewidth}{
\includegraphics[width=0.99\linewidth]{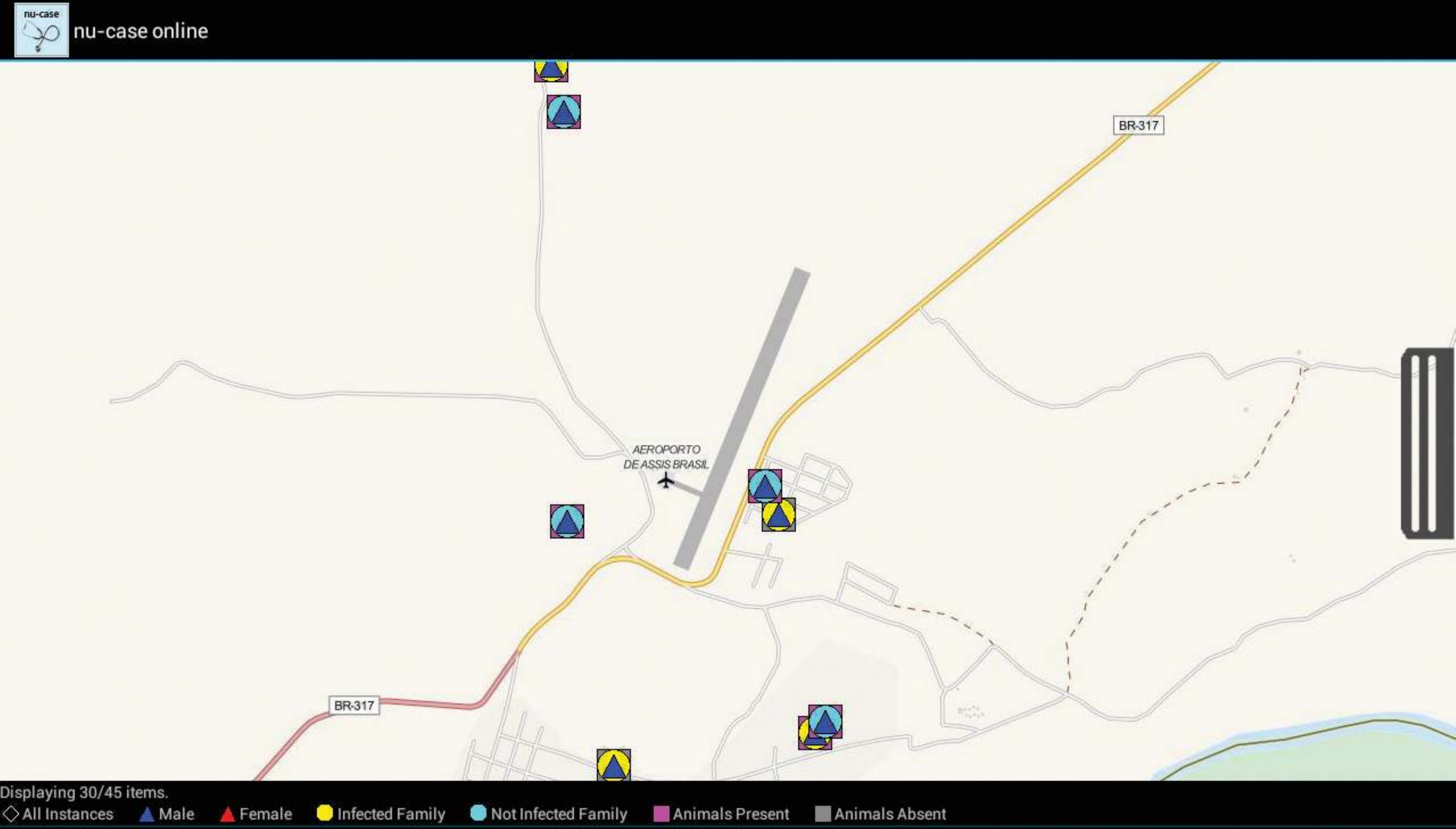} d)}
\caption{The group visualization changes when individual users
share more details (a), or move their view (b), and the corresponding
personal visualizations shown on their tablet display (c, d).
}
\label{fig:first-device-visualization}
\end{figure*}

Figure~\ref{fig:first-device-visualization} shows examples of these group (top) and individual (bottom) modes, simultaneously on a tablet device and a large computer display. Individual analysts were able to share more detailed views of their personal visualizations, such as those resulting from their visual analysis in decoupled interaction mode. 

The images on the left-hand side of Figure~\ref{fig:first-device-visualization} show a selected set of patient cases with minimal case details (using black squares), while the right-hand side images illustrate how more specific details of cases are shared in another part of the shared visualization (using various icons created based on individual case attributes). Cases that are not in areas being viewed by individual analysts, or selected by them to be shared, remain invisible on the group visualization. Visualizations shared with the group by individual researchers are overlaid on the group visualization as separate overlays, displaying their individual areas of view, selected cases, level of details, and other individual configuration parameters that specify the individual displays. Personal views are identified on the large screen as semi-transparent  rectangular regions and their geometries vary according with the actual size, screen resolution, and zoom level of the individual devices connected to the large screen.

Our observations of group meetings prior to the development of the system revealed that most disciplines relied on maps to contextualise the data presented at the meetings. This was clearer in the work of epidemiologists, who presented statistical data on a large display, using a slides presentation software, but made frequent references to a map of the region of interest  by switching back and forth between slides. Although analyses of patient case data sets were often presented, the presentation of these cases was not well supported by maps, which would have required preparation of individual maps well in advance of the meeting. In the absence of such maps, presenters had to rely on textual tables and verbal references to regions of interest, and information had to be essentially exchanged through verbal explanations.  Different kinds of data were presented during these meetings, including disease occurrence statistics and their geographical distribution, data on the utilisation of public healthcare services, and public policy documents, shown for purposes of strategic planning, assessment of interventions and identification of areas for future interventions. 

Following the initial experience of interacting with the system, users suggested many possible enhancements. These suggestions included using the large screen for collaborative creation of tables, sharing of geo-referenced photographs, support for integration of externally linked data and overlays (e.g., an overlay layer showing the ethnic composition of the population in a given area of interest), and ability to annotate the shared area by using the mobile devices for documentation of meeting decisions and outcomes.

\section{Conclusions}

In this paper, we have reviewed the application of interactive maps to support complex analytical tasks in epidemiology and public health, highlighting some of the challenges that remain in this area. We illustrated map-based interfaces that addressed some of these challenges, and reported the feedback received from a multidisciplinary group of researchers, health care professionals, and public health managers who regularly made use of maps in their work.

While much progress has been made towards integrating spatial and temporal aspects of epidemiological work into usable map-based interfaces, research is still needed on issues relating to spatial data granularity, temporal consistency, privacy, integration of data from diverse sources, adequacy of the underlying epidemiological models to interactive maps, and improved support for causal inference.


\balance

\bibliographystyle{ACM-Reference-Format}
\bibliography{mapii2022}

\end{document}